# Coupling of the wings and the body dynamics enhances damselfly maneuverability

Samane Zeyghami, Haibo Dong

Mechanical and Aerospace Engineering, University of Virginia, Charlottesville, VA, 22904

In flapping flight, motion of the wings through the air generates the majority of the force and torque that controls the body motion. On the other hand, it is not clear how much effect the body motion imposes on the wings. We investigated this connection via analyzing fast yaw turns of three different species of damselfly. In this combined experimental and theoretical study, we show that the dynamics of the wings and the body are coupled together in low frequency flapping flight. As a result, damselflies benefit from a passive mechanism for enhancing the bilateral wing pitch angle asymmetry to sustain the body rotation. A physics-based model derived from this mechanism is proved valid for linking morphology, kinematics and dynamics of the wing and the body of the flying insects in fast turning maneuvers.

*Introduction.* -In order to survive, flying insects not only need to stay aloft but also need to steer and maneuver in the air. Insects use a variety of different kinematics alternations to perform an aerial maneuver such as change in the wing stroke amplitude [1], wing pitch angle [2], or speed and timing of wing rotation [3,4]. Flight muscles in the thorax are responsible for implementing the fine adjustments to the wing kinematics [5]. Although insects are capable of actively controlling the wing kinematics, the mechanics of the wing and its hinge allows the wing to passively flip in response to the aerodynamic, inertial and the elastic torques. There is a growing body of evidence that supports passive maintenance of the wing's angle of attack [6] as well as the passive wing rotation at stroke reversal [7,8] in hovering flight. In addition, it was suggested that fruit flies combine the active and the passive wing kinematic changes in order to perform a turn maneuver [9]. Yet, the majority of the previous studies on the wing kinematics control and its association with flight dynamics were focused on high flapping frequency insects such as fruit flies. However, insects such as damselflies move their wings through the air with a significantly lower velocity. In fact, the neuromuscular structure that drives the wing and controls its motion is essentially different for these insects and there is much to be learned about the similarities and differences in the ways these insects use their wings for controlling the body motion.

One of the fundamental differences between the flight dynamics of low and high frequency flapping fliers is that in low frequency flapping flight, the time scales of the wings and the body motions are close in magnitude. This is not the case in high flapping frequencies. Therefore (in high frequency flapping flight) the dynamics of the wings and the body are (assumed to be) decoupled and the effect of the body motion is only considered on changing the aerodynamic force and torque that act on the body [10-12]. Combining a theoretical and experimental approach, here we argue that in low frequency flapping flight, the change in the aerodynamic force and torque not only alters the body motion but also affects the motion of the wings. To gain insight into the potential effect that the body motion can impose on the wing motion, we have shot more than 20 videos of free flight yaw turns of damselflies ranging between 20 to 170 degrees. Morphology and kinematics parameters of a representative turn of each species is presented in Table 1. An accurate 3D surface reconstruction technique is then used to extract the three dimensional body and wing kinematics [13]. The extracted flight data reveals that the rotational velocity of the damselfly's body is about 15-30% of that of its wings. At such speeds, the incoming velocity to the wings changes significantly due to the body rotation and the wing motion is no longer independent from the body motion.

Table 1. Morphology and kinematics of the three damselfly species in this experiment.

| | Species | Body weight (mg) | Body length (mm) | Forewing span (mm) | Hindwing span (mm) | Forewing chord (mm) | Hindwing chord (mm) | Flapping frequency (Hz) | Flapping amplitude (deg) |
|---|---|---|---|---|---|---|---|---|---|
| **Damselfly #1** | *Hetaerina Americana* | 85 | 42 | 29 | 28 | 6 | 6 | 25 | 70 |
| **Damselfly #2** | *Argia Apicalis* | 31 | 36 | 23 | 22 | 4 | 4 | 25 | 100 |
| **Damselfly #3** | *Ebony Jewelwing* | 64 | 44 | 27 | 26 | 9 | 10 | 17 | 60 |

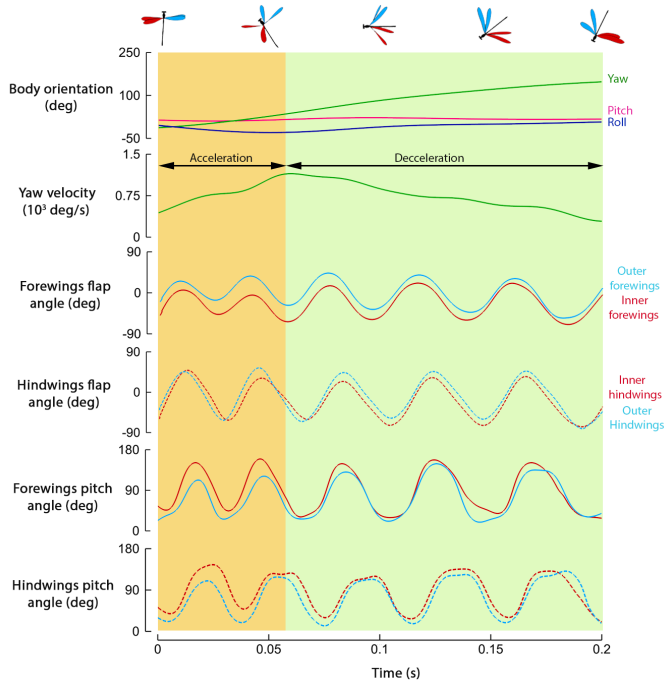

**Fig. 1**. The body position and orientation as well as the wings flap and pitch angles. A top view of the damselfly on the top of the figure visualizes the yaw throughout the maneuver. The flap angle is the angle that wing travels in the stroke plane and pitch angle is the angle between the wing's mid chord line and the stroke plane. The accuracy of the body yaw, pitch and roll angle measurements are ±1, ±2, ±5 deg. The measurement of the wing flapping angle is accurate up to ±1 deg. The accuracy of the wing pitch angle measurements is ±2 deg.

*Damselfly's yaw turn dynamics.*- The three representative turns range between 160 to 170 degrees (Table 2). All maneuvers are very close to a pure yaw turn. The body and the wing kinematics of the damselfly #1 is shown in Fig. 1. Consistently for all three damselflies, the body yaw velocity rises during the first two wingbeats. This phase is referred to as acceleration phase in Fig. 1. The angular velocity of the body then started to drop slowly. This phase which is referred to as the deceleration phase lasts for about 3 wingbeats. The majority of the heading change takes place during this phase. The maximum yaw velocity as well as the flapping frequency varies between the species (Table 2). Yet, in the deceleration phase, the half-stroke-averaged (averaged over a downstroke or an upstroke) ratio of the body to the wing angular velocity, $\Omega/\omega$, averaged at about $0.2 \pm 0.085$.

To execute a yaw turn, in downstroke, the damselfly flaps the inner wings with a higher pitch angle compared to the outer wings (Fig. 1). That causes the drag to increase on the inner wings and turn the body. In upstroke, thrust increases on the outer wings due to the higher wing pitch angle which causes the body to turn in the same direction. The high magnitude of the bilateral wing pitch asymmetry within the first two wingbeats causes the body rotational velocity to rise rapidly. The magnitude of the wing pitch asymmetry drops for the deceleration phase. It was previously shown that the decelerating and stopping the turn does not require significant help from the active control. The asymmetric variation in the incoming velocity of the bilateral wings causes a strong damping effect known as Flapping Counter Torque (FCT) which is expected to quickly diminish the angular velocity of the damselfly body [11]. FCT predicts the yaw velocity of the damselfly to fall to half its value in less than a wingbeat. However, this is not consistent with our observation. All the damselflies in our experiment were able to maintain the body velocity for a significantly longer time.

To probe the dynamics of the deceleration phase, we investigated the association between the wing and the body motion. Fig. 2 shows the half-stroke-averaged wing pitch angle versus the corrected wing angular velocity (more statistical data are provided in *SM*). A strong correlation is found between these two parameters for both forewings and hindwings. The corrected wing angular velocity is defined as the flapping velocity of the wing, with respect to the body, corrected by the rotational velocity of the body. For instance, in downstroke, the corrected angular velocity of the outer wing is the sum of the flapping velocity and the body yaw velocity. The vector of the flapping velocity of the wing is slightly tilted with respect to the vector of the body yaw velocity (less than 20 deg). This is because of the inclination of the wing stroke planes. The slope of the regression line between the half-stroke-average wing pitch angle and the corrected angular velocity of the wings is very close for the forewings and the hindwings but the hindwings move through the stroke plane with a higher pitch angle. Due to the differences in the morphology and kinematics, this slope varies among the species (*SM*). Although the statistical correlation observed in the flight data does not necessarily imply causation, it suggests existence of a relationship between the wing pitch angle and the corrected wing angular velocity. To prove that the rotation of the body and the consequent change in the incoming velocity to the wings has in fact caused the variations in the wing pitch angles, we construct a physics-based model of the wing pitching and show that the corrected wing angular velocity can successfully predict the measured wing pitch angles.

*Response of the physics-based model of the wing to the body rotation.*- Insect wings are flexible and deform under the action of the aerodynamic and inertial forces [14,15]. However, for simplicity, we modeled the wing as a rigid wing with an elastic pitching hinge [6,7,15-17]. The flapping motion of the wing through the stroke plane is prescribed. The dynamics of the wing pitching is governed by the Eq. (1).

Table 2. Kinematics of the body yaw maneuver.

|  | Turn amplitude (deg) | Maneuver duration (ms) | Maneuver duration (wingbeats) | Max yaw velocity ($10^3$ deg/s) |
|---|---|---|---|---|
| **Damselfly #1** | 160 | 200 | 5.5 | 1.2 |
| **Damselfly #2** | 170 | 180 | 5.0 | 1.7 |
| **Damselfly #3** | 170 | 300 | 5.0 | 0.7 |

$$I_{xx}\dot{\omega}_x + I_{xy}(\dot{\omega}_y - \omega_x\omega_z) = \tau_{aero} + \tau_{elastic} \quad (1)$$

$(\omega_x, \omega_y, \omega_z)$ is the vector of the angular velocity of the wing in the wing coordinate system. *x*-axis of the wing lies on the rotation axis. *y*-axis points forward and *z*-axis is in the direction of the vector product of *x* and *y* axes. $\tau_{aero}$ and $\tau_{elastic}$, respectively, are the aerodynamic and elastic reaction torques about the wing's axis of rotation. $I_{ij}$ is the tensor of moment of inertia of the wing in the wing coordinate system. It is calculated assuming homogeneous mass distribution across the wing area. The elastic torque about the wing axis of rotation is linearly proportional to the wing deflection. The pitch angle of the wing is determined by the balance between the inertial, aerodynamic and elastic reaction forces, with inertial forces playing a less important role [6,18]. *Ch* number defines the relative ratio of the fluid pressure to the elastic force [6].

$$Ch = \frac{\rho_f V_w \bar{c}^4 f}{G} \quad (2)$$

$\rho_f$ is the fluid density. $\bar{c}$ and $f$ are wing's average chord and flapping frequency, respectively. $V_w$ is the average translational velocity of the wing tip.

When an insect experiences a whole body rotation, in downstroke (upstroke), the net incoming velocity increases on the outer (inner) wing and it decreases on the inner (outer) wing. This results in asymmetric variation in the fluid dynamic pressure on the wings. Since, in the absence of active control, the elastic properties of the wing hinge remain unchanged, the relative strength of the fluid force (and torque) compared to the elastic reaction force, *Ch* number, varies. The asymmetric change in *Ch* number then causes an asymmetric response of the wings' pitch angles. For instance, on the outer wing during downstroke, the fluid forces are strengthened which (since the center of pressure of the wing is located behind the axis of rotation) results in a stronger pitch down torque being exerted on the wing (Fig. 3a). This is consistent with the experimental relationship we have observed between the corrected wing angular velocity and the wing pitch angle (in Fig 2). However, the effectiveness of this phenomenon depends on the extent to which *Ch* number changes due to the body rotation. The relative change in *Ch* number of the bilateral wings depends on the relative magnitude of the angular velocity of the body compared to that of the wing, $\Omega/\omega$.

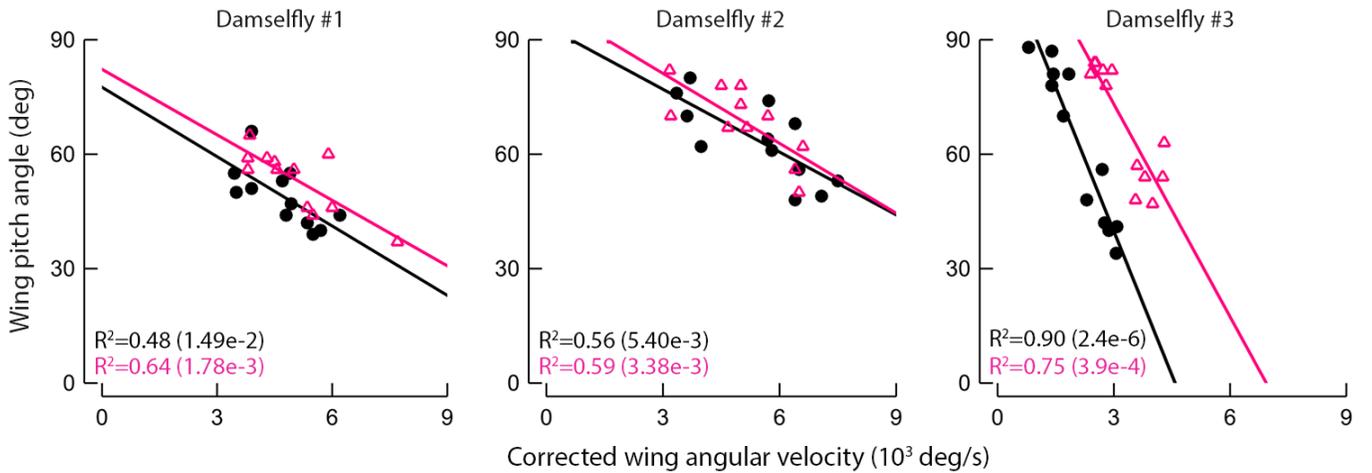

**Fig. 2.** Half-stroke-averaged wing pitch angle versus the corrected wing angular velocity during the deceleration phase. The black and deep pink, respectively, represent the forewings and hindwings. On each graph, the correlation coefficients, $R^2$, are shown with the same color as the respective regression line. P-values are shown in the brackets. The difference in the least square regression slop suggests difference in *Ch* for these insects.

$$Ch_{maneuver} = \frac{\rho_f L_w \bar{c}^4 f}{G}\omega(1\pm\frac{\Omega}{\omega}) \qquad (3)$$

For a damselfly in a fast yaw turn, *Ch* of the wings may vary up to 35% compared to its value when the body rotational velocity is zero. Thus, we hypothesize that the rotation of the body and the consequent asymmetric change in *Ch* number of the bilateral wings causes asymmetry in their dynamics.

To test this hypothesis, we conduct a numerical experiment. In this experiment, the pitch angles of the bilateral wings are numerically calculated by integrating Eq. (1). The geometry, the flapping kinematics and the hinge torsional stiffness of the bilateral wings are kept identical. The rotation of the body is prescribed by a second order polynomial with a peak value of 1500 deg/s (Fig.3b). At the beginning of the maneuver, the two wings have an identical pitch angle. As the body angular velocity increases, the pitch angles of the wings start to depart from each other. We termed this phenomenon as Passive wing Pitch Asymmetry (PPA) because the alternations in the wing kinematics are not actively forced by the flight muscles in the thorax but are passively induced by the body rotation. The passive variations in the wing pitch angles are in such way that it helps maintain the yaw velocity of the body. Our results suggest that even if the active modulation of the wing kinematics stays uniform throughout the maneuver, this passive mechanism will augment the asymmetry in a similar fashion to what was observed in the insect maneuvers [19,20].

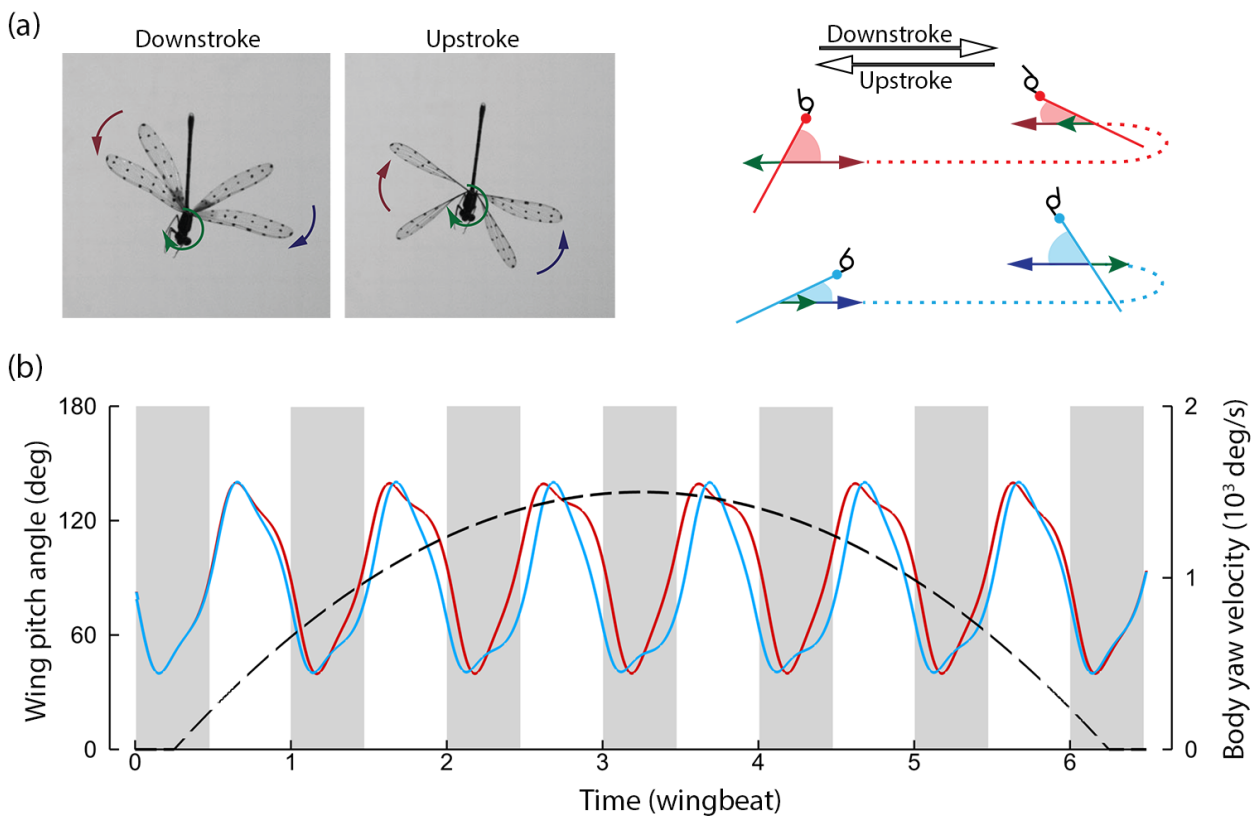

**Fig. 3.** (a) Damselfly in a yaw turn to the left. The angular velocity of the body enhances the velocity of the outer wing in downstroke and reduces it in upstroke (vice versa for the inner wing). Increasing or decreasing the velocity changes the balance between the aerodynamics and elastic torque and consequently causes asymmetric alternation in the bilateral wing pitch angles. The outer and inner wings are shown by blue and red, respectively. The body rotation is shown in dark green. (b) Stroke by stroke augmentation of the passive wing pitch asymmetry during a model turn maneuver. The red and blue lines represent the pitch angles of the inner and outer wings, respectively. The black line shows the time course of the prescribed body angular velocity which is shown on the right vertical axis. The body turn velocity is prescribed by a second order polynomial. Downstrokes are shaded gray.

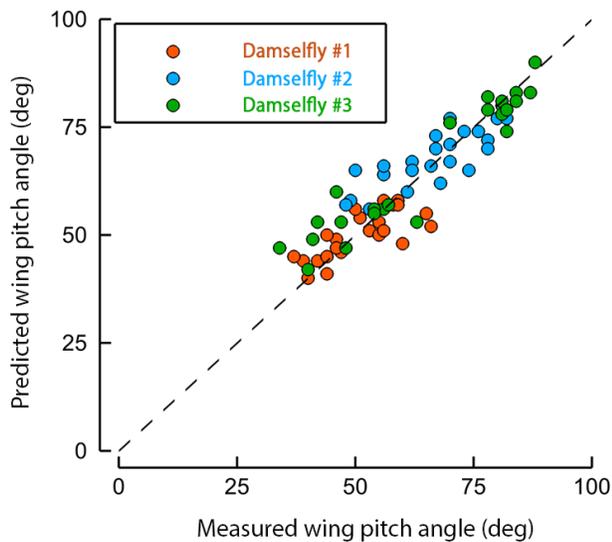

**Fig. 4.** Predicted versus the measured wing pitch angles for all three damselflies. The predicted values are calculated by matching the half-stroke-averaged wing flapping velocity as well as the body yaw velocity to the experimentally measured data.

*PPA in damselfly's yaw turn.-* Fig. 4 shows the predicted versus the measured half-stroke-averaged wing pitch angle for all three insects (during the deceleration phase). In the calculation of the wing pitch angle we matched the half-stroke averaged flapping velocity and body yaw velocity to the experimentally measured values. It is difficult to measure the wing hinge stiffness experimentally. However, this parameter can be calculated if $C_h$ as well as the wing geometry and kinematics are known (Eq. (2)). For each damselfly, $C_h$ of the wings is estimated from the experimentally observed slope of the wing pitch angle versus the corrected wing velocity (*SM*). Note that in Fig. 3b the hinge stiffness of the fore and the hind wings is not necessarily the same. However, we assume that the right and the left wings (of each pair) have identical values. In addition, this value remains constant throughout the deceleration phase (assuming there is no active control). The predicted pitch angle values are in excellent agreement with the measurement regardless of the difference in the morphology and kinematics of the wings and the body among species. Therefore, we infer that the rotation of the damselfly's body induces passive alternations in the wing pitch angles by shifting the balance between the aerodynamics and the elastic torque exerted about the wing's rotation axis. This variation is bilaterally asymmetric and increases due to an increase in the body rotational velocity. The asymmetric variation in the bilateral wing pitch angles arises from the difference in the velocity of the outer and the inner wing and therefore is restricted to flapping flight. As a result a damselfly is capable of maintaining its body rotational velocity for a longer time (after the initial acceleration phase) even if no further active control is employed. For instance, the yaw velocity half-life of the damselfly #1 is measured to be about 2.5 wingbeats, which is significantly longer than the prediction by FCT (about 0.77 wingbeat). This suggests that these insects might have adapted other passive mechanisms, such as PPA, to overcome the damping effects and enhance maneuverability.

*Conclusions.*-The current work reveals the critical importance of considering the effect of the body motion on modulating the wing motion in damselfly flight. Further studies will examine the existence of similar effects in maneuvering flight of other low flapping frequency insects such as dragonflies. Our results show that the coupling between the wing and the body dynamics gives rise to passive alternation in the bilateral wing pitch angles when the insect body is engaged in a fast yaw turn. Specially, we showed that a damselfly predominantly uses this passive mechanism for maintaining the bilateral wing pitch asymmetry that generates the aerodynamic yaw torque for keeping the body in rotation.

Taken together, these findings emphasize the important role that the aerodynamics of flapping flight plays in the wing kinematics control. Previously, many studies investigated the effect of the wing motion on generating the unsteady aerodynamic forces in insect flight [21-25]. However, there is much to be learned about the reverse relationship. Wings of the insects constantly interact with the surrounding air, even though they are powered from within the thorax by the flight muscles. Here, we showed that the combination of the wing biomechanics and the aerodynamics of flapping flight allow modulations of the wing motion without additional assistance from the flight muscles in the thorax. Yet, more studies are necessary to unravel the mutual relationship between these two aspects of flapping flight.

**ACKNOWLEDGMENTS.** We thank Emma Lois Mitchell and Wen Zhang for 3D surface reconstruction of the damselfly flights. The authors gratefully acknowledge support from National Science Foundation [grant number CEBT-1313217] and Air Force Research Laboratory [grant number FA9550-12-1-007] monitored by Dr. Douglas Smith.

# Supplementary Materials for "Coupling of the wings and the body dynamics enhances damselfly maneuverability"

Samane Zeyghami, Haibo Dong

Mechanical and Aerospace Engineering, University of Virginia, Charlottesville, VA, 22904

*Videography and motion tracking.*- The turn flights of the damselflies are recorded by three synchronized Photron FASTCAM SA3 60K high-speed cameras with 1024×1024 pixel resolution. They are aligned orthogonally to each other on an optical table and operate at 1000 Hz with at least a 1/20000 second shutter speed. The 3D surface reconstruction technique is then applied to the raw output from the high speed cameras. Details about this technique and its accuracy can be found in [1]. Although the body of the damselfly stays relatively rigid, the wings deform when flapping. The rigid wing kinematics is measured by defining a rigid plane made by a reference span line, created by connecting the wing root and the tip, and a reference chord line. This rigid plane defines the wing orientation with respect to the body. The orientation of the wing is specified by flapping angle, which is the angle that the wing travels in the stroke plane, and the pitch angle which indicates the angle between the mid chord line and the stroke plane. The instantaneous deviation of the wing from the stroke plane is determined by the wing deviation angle. The variation of the deviation angle is small compared to the flapping and pitching angles. The stroke plane is defined as the mean square plane of the position of the wing marker points during one wingbeat. The orientation of this plane with respect to the horizontal plane is the stroke plane angle.

*Statistical analysis of the flight data.*- To probe the dynamics of the deceleration phase of the damselfly yaw turn maneuver, we investigated the link between the wing and the body motion. Our measurements show that the half-stroke-averaged wing pitch angles as well as the bilateral pitch angle difference are independent from the body angular velocity (Table 1). In most insects, the central control system (which controls the wing motion) receives feedback from the body motion [2]. Therefore, the lack of correlation between the body motion and the wing pitch angle suggests that the measured wing pitch values are not dominantly forced by the active control. In addition, the relationship between the wing pitch angle and the flapping velocity is somewhat random (Table 1). However, a strong correlation was found between the wing pitch angle and the net incoming velocity to the wing (main text).

Linear correlation analysis is performed on all the experimental data. All the statistical analysis is done in MATLAB.

*Physics-based model of the wing pitching.*- Each damselfly's wing is digitized from its picture. We normalize the lengths along the wing span and chord by wing length, $L_w$, and maximum wing chord length, $C_{max}$, respectively. A 4$^{th}$ order polynomial curve is then fitted to the leading and trailing edges of the wing, separately. The axis of rotation always passes through the origin of the wing coordinate system (Fig. 1a). The orientation of the wing is specified by flapping and pitch angles which are defined in a similar fashion for the real damselfly flight data. In all the numerical experiments, the deviation of the wing from the stroke plane is assumed zero and the flapping angle is modeled by a sinusoidal function. The body yaw axis is normal to the wing stroke plane.

The flapping motion of the wing is prescribed and the pitching is governed by the interaction of the inertial, elastic and the aerodynamic torque (Eq. (1) in the main text). The governing equation of the pitching dynamics is solved in a frame attached to the wing which its x-axis is aligned by the wing's rotation axis (Fig. 1A). The elastic torque about the wing rotation axis is modeled by a linear torsion spring.

$$\tau_{elastic} = -G(\psi - \psi_0) \qquad (1)$$

$\psi_0$ is the spring rest angle and was set to 90$^o$, showing the tendency to keep the wing vertical.

Table 1. Correlation coefficients

|  | Pitch angle vs. flapping velocity | | Pitch angle vs. body yaw velocity | | Bilateral pitch angle difference vs. body yaw velocity | |
| --- | --- | --- | --- | --- | --- | --- |
|  | Forewings | Hindwings | Forewings | Hindwings | Forewings | Hindwings |
| **Damselfly #1** | 0.079 | 0.252 | 0.018 | 0.006 | 0.056 | 0.138 |
| **Damselfly #2** | 0.390 | 0.009 | 0.054 | 0.110 | .0880 | 0.347 |
| **Damselfly #3** | 0.094 | 0.056 | 0.007 | 0.028 | 0.622 | 0.054 |

*Wing hinge stiffness estimation.*- We were not able to directly measure the wing hinge stiffness but that value can be calculated once the wing and the body kinematics, wing geometric parameters and the wing pitch angles are measured. Our model predicts that the slope of the variations of the wing pitch angle versus the corrected wing velocity is a function of $Ch$ and thus the hinge stiffness. Note that G is the only free parameter (in the expression for $Ch$) once the wing geometry and kinematics are measured. Changing the body rotational velocity (in steps of 250 deg/s) we calculate the average wing pitch angle and plot the results. We then vary G and repeat the same calculation. This plot is shown for a forewing of the damselfly #3 if Fig. 1b. As was expected, by decreasing G (and therefore increasing $Ch$), the slope of the variation of the wing pitch angle with the corrected wing velocity increases. No attempt was made to find the exact value of G. Yet, the G value that gives the closest slope to the experimental flight data (Fig. 2 in the main text) was used for the wing pitch angle prediction. The fact that the G value attained this way gives an accurate prediction of the wing pitch angle further proves the consistency of our theory with the observed flight behavior.

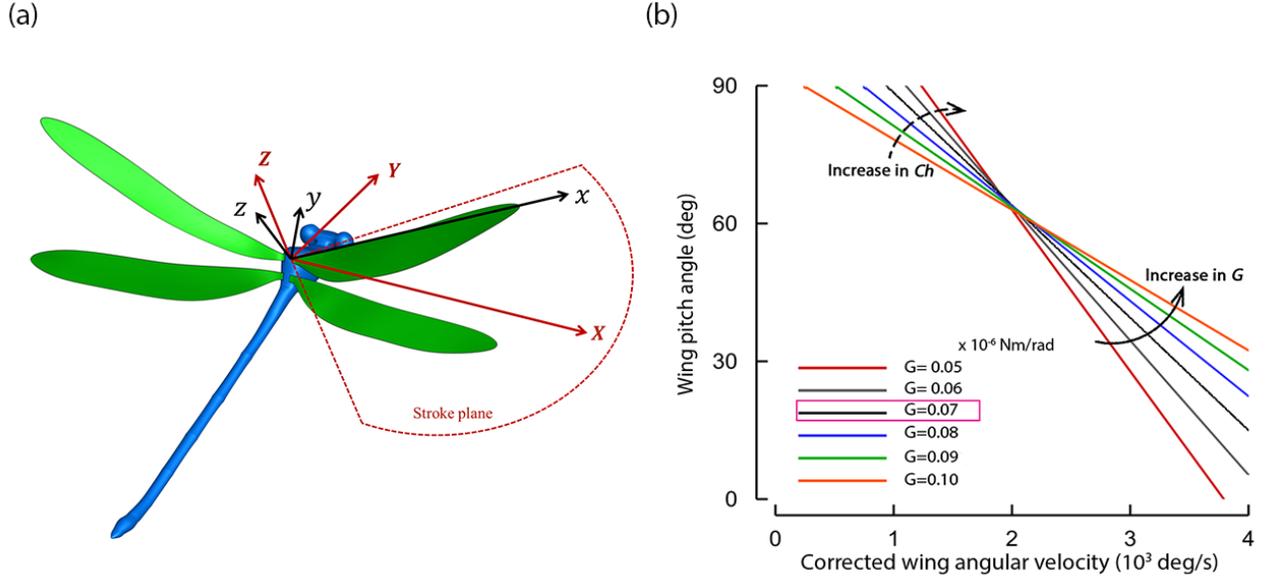

**Fig. 1.** (a) Coordinate systems of the wing and the body. (b) The wing pitch angle versus the corrected wing velocity for a wing with the geometry and kinematics of the damselfly #3. The hinge stiffness was varied until the slope of the relationship resembles that of the experimental observation. This was achieved for G=0.07 μNm/rad (the theoretical and experimental slopes are 24.49 and 25.02, respectively. Similar procedure was performed for forewings and hindwing of each damselfly.

*Quasi-steady prediction of the aerodynamic pitching torque.*-The aerodynamics pitching torque about the wing rotation axis is due to the deviation of the center of action of the translational, rotational and the added mass force from the wing's pitching axis. There is also a net couple due to the fluid added mass as well as a net (rotational) circulatory torque which is proportional to the product of the translational and rotational velocities of the wing [3]. The rotational circulatory torque does not have an equivalent partner in the force description. That is because it is a result of the action of two opposing force components with equal strengths but different centers of actions [3,4]. The dissipative torque component originates form the action of the translational drag force as well as the resistance of the air to the wing rotation. The expression of the aerodynamic pitching torque on a 2D slice of the wing is given in Eq. (2). The total aerodynamic pitching torque is calculated by integrating $d\tau_{aero}$ over the wing length.

$$d\tau_{aero} = d\tau_{am} + d\tau_{trns} + d\tau_{rot} + d\tau_{diss}$$
$$d\tau_{am} = -I_a \dot{\omega}_x + \frac{1}{4}\pi\rho_f c^2 \left(c_i/2 - x_i\right)(d\upsilon_y/dt)dr$$
$$d\tau_{trns} = \rho_f \Gamma_{trns} \upsilon_x (x_i - x_{Cp_i}) dr \quad (2)$$
$$d\tau_{rot} = -\rho_f \Gamma_{rot} \upsilon_x (x_{Cp_i} - x_i) - \frac{1}{2}\rho_f \Gamma_{rot} \upsilon_x (0.75c_i - x_i)dr$$
$$d\tau_{diss} = dF_y^v (x_i - x_{Cp_i}) - \frac{1}{8}\rho_f C_{rd} |\omega_x|\omega_x \left(y_{LE_i}\left|y_{LE_i}\right|^3 - y_{TE_i}\left|y_{TE_i}\right|^3\right)dr$$

Indices *am*, *trns*, *rot* and *diss* stand for the contributions of the added mass, translational, rotational and dissipative torques, respectively. $C_{rd}$ is the rotational damping coefficient which was set to 5 [5]. $c_i$ is the chord length of the $i^{th}$ cross section of the wing along the span and $x_i$ is the distance of the rotation axis from the section's leading edge. $y_{LE_i}$ and $y_{TE_i}$ are the distance of the leading and the trailing edges of the $i^{th}$ cross section measured from the pitching axis. $v_y$ and $v_z$ are the velocity components of the wing parallel and normal to the wing. $x_{cp_i}$ represents the location of the center of pressure and was estimated by Eq. (3) [6]. Assuming the 2D wing section has negligible thickness, the added mass coefficients can be expressed as follows:

$$I_a = \frac{1}{4}\pi \rho_f c_i^2 \left[\frac{1}{32}c_i^2 + (c_i/2 - x_i)^2\right] \quad (3)$$

The torque due to the added mass has two components: rotational, which is proportional to the angular acceleration, and translation, which results from the deviation of the rotation axis from the geometric center. The added mass is an unsteady aerodynamic effect which can potentially have significant contributions to the force and torque generation in flapping flight [7].

The translational circulation has the same form as in [8]. The rotational circulation is a function of the location of the rotation axis.

$$\Gamma_{rot} = c_r(.75 - \hat{x}_0)\dot{\psi}c^2 \quad (4)$$

$\hat{x}_0$ is the dimensionless distance of the rotation axis measured from the leading edge in chord lengths. $c_r$ is a constant which is set to $\pi$. It was previously proved that this form of the rotational circulation shows good accuracy in describing the aerodynamics of flapping flight [9,10].

The validity of the aerodynamic torque model was tested by comparing the predicted passive wing pitch angle with the measurements in three experiments presented in ref. [5]. The results are shown in Fig. 2a. The aerodynamic torque predicted by Eq. (2) is plotted versus the experimental torque as well (Fig. 2b). The experimental torque was not directly reported in that reference. Here, it is calculated using the experimentally measured wing orientation by balancing the dynamics of the wing by inertial, aerodynamic and elastic torques (Eq. (1) in the main text).

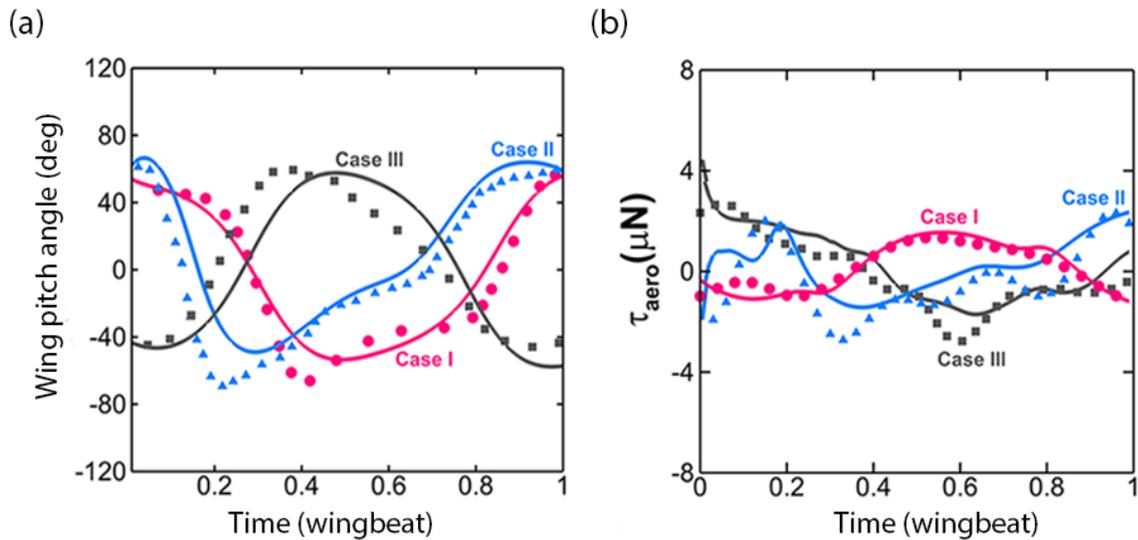

**Fig. 2.** Prediction of the passive wing pitch angle (a) and aerodynamic pitching torque (b) for the crane fly model wing flapping at 100 Hz. Our results (solid lines) are compared to three experiments in ref. [5] (symbols).